\documentclass[twocolumn,trackchanges]{aastex62}

\usepackage{acronym}
\usepackage{graphicx}
\usepackage{amsmath,amsfonts,amssymb}
\usepackage{mathrsfs}
\usepackage[utf8]{inputenc}
\usepackage[normalem]{ulem}

\newcommand{\ie}{\textit{i.e.}}

\usepackage{color}

\newcommand{\fornax}{F\textsc{ornax}}

\acrodef{ADM}{Arnowitt-Deser-Misner}
\acrodef{AMR}{adaptive mesh-refinement}
\acrodef{BH}{black hole}
\acrodef{BBH}{binary black-hole}
\acrodef{BHNS}{black-hole neutron-star}
\acrodef{BNS}{binary neutron star}
\acrodef{CCSN}{core-collapse supernova}
\acrodefplural{CCSN}[CCSNe]{core-collapse supernovae}
\acrodef{CMA}{consistent multi-fluid advection}
\acrodef{CFL}{Courant-Friedrichs-Lewy}
\acrodef{DG}{discontinuous Galerkin}
\acrodef{HMNS}{hypermassive neutron star}
\acrodef{EM}{electromagnetic}
\acrodef{ET}{Einstein Telescope}
\acrodef{EOB}{effective-one-body}
\acrodef{EOS}{equation of state}
\acrodef{FF}{fitting factor}
\acrodef{GR}{general-relativistic}
\acrodef{GRLES}{general-relativistic large-eddy simulation}
\acrodef{GRHD}{general-relativistic hydrodynamics}
\acrodef{GRMHD}{general-relativistic magnetohydrodynamics}
\acrodef{GW}{gravitational wave}
\acrodef{ILES}{implicit large-eddy simulations}
\acrodef{LIA}{linear interaction analysis}
\acrodef{LES}{large-eddy simulation}
\acrodefplural{LES}[LES]{large-eddy simulations}
\acrodef{MRI}{magnetorotational instability}
\acrodef{NR}{numerical relativity}
\acrodef{NS}{neutron star}
\acrodef{PNS}{protoneutron star}
\acrodef{SASI}{standing accretion shock instability}
\acrodef{SGRB}{short $\gamma$-ray burst}
\acrodef{SPH}{smoothed particle hydrodynamics}
\acrodef{SN}{supernova}
\acrodefplural{SN}[SNe]{supernovae}
\acrodef{SNR}{signal-to-noise ratio}
\acrodef{ZAMS}{zero age main sequence}

\shorttitle{Gravitational Waves from Core-Collapse Supernovae}
\shortauthors{Radice et al.}

\begin{document}

\title{Characterizing the Gravitational Wave Signal from Core-Collapse
Supernovae}
\correspondingauthor{David Radice}
\email{dradice@astro.princeton.edu}

\author[0000-0001-6982-1008]{David Radice}
\affiliation{Institute for Advanced Study, 1 Einstein Drive, Princeton, NJ 08540, USA}
\affil{Department of Astrophysical Sciences, Princeton, NJ 08544, USA}
\author{Viktoriya Morozova}
\affiliation{Department of Astrophysical Sciences, Princeton, NJ 08544, USA}
\author{Adam Burrows}
\affiliation{Department of Astrophysical Sciences, Princeton, NJ 08544, USA}
\author{David Vartanyan}
\affiliation{Department of Astrophysical Sciences, Princeton, NJ 08544, USA}
\author{Hiroki Nagakura}
\affiliation{Department of Astrophysical Sciences, Princeton, NJ 08544, USA}

\begin{abstract}
We study the gravitational wave signal from eight new 3D core-collapse
supernova simulations. We show that the signal is dominated by $f$- and
$g$-mode oscillations of the protoneutron star and its frequency
evolution encodes the contraction rate of the latter, which, in turn, is
known to depend on the star's mass, on the equation of state, and on
transport properties in warm nuclear matter. A lower-frequency component
of the signal, associated with the standing accretion shock instability,
is found in only one of our models.  Finally, we show that the energy
radiated in gravitational waves is proportional to the amount of
turbulent energy accreted by the protoneutron star.
\end{abstract}

\section{Introduction.}
Core-collapse supernovae \acused{CCSN} (\ac{CCSN})\acused{SN} have long
been considered promising sources of \acp{GW} for ground-based detectors
\citep{Wheeler:1966tg, Finn:1990qf, Ott:2008wt, Kotake:2011yv} such as
Advanced~LIGO \cite{TheLIGOScientific:2014jea}, Advanced~Virgo
\citep{TheVirgo:2014hva}, and KAGRA \citep{Aso:2013eba}. The combined
observation of \acp{GW}, neutrinos, and photons \citep{Nakamura:2016kkl}
from the next Galactic \ac{CCSN} could unveil the mechanism by which
massive stars explode at the end of their lives, resolving a puzzle that
has eluded the scientific community for more than 50 years
\citep{Burrows:2012ew, Janka:2012wk, Muller:2016izw}. Multi-messenger
observations of the next galactic \ac{CCSN} could also constrain the
properties of matter at extreme densities and the interior structure of
massive stars, and reveal the origin of many of the chemical elements.

The current understanding of the \ac{GW} signal from \ac{CCSN} is for
the most part derived from the analysis of 2D (axisymmetric) simulations
\citep{Finn:1990qf, Dimmelmeier:2002bm, Shibata:2004nv,
Dimmelmeier:2007ui, Marek:2008qi, Murphy:2009dx, 2009ApJ...697L.133K,
Mueller:2012sv, Cerda-Duran:2013swa, Abdikamalov:2013sta,
Yakunin:2015wra, Pan:2017tpk, Morozova:2018glm}, or 3D simulations with
simplified microphysics \citep{1997A&A...317..140M, Rampp:1997em,
Fryer:2004wi, Shibata:2004kb, Ott:2006eu, Ott:2010gv, Ott:2012mr,
Kuroda:2013rga, Kuroda:2016bjd, Hayama:2016kmv,  Kuroda:2017trn,
Hayama:2018zgb, OConnor:2018tuw, Powell:2018isq}. However, a number of
sophisticated neutrino-radiation-hydrodynamics simulations of \ac{CCSN}
have become available in the past several years \citep{Tamborra:2013laa,
Melson:2015tia, Lentz:2015nxa, Melson:2015spa, Roberts:2016lzn,
Muller:2017hht, Ott:2017kxl, Summa:2017wxq, Kuroda:2018gqq,
OConnor:2018tuw, Vartanyan:2018iah, Glas:2018oyz}. Gravitational wave
signals have been published for ten of these models
\citep{Andresen:2016pdt, Yakunin:2017tus, Andresen:2018aom}. However,
even though these simulations exhibit some common qualitative features,
it is difficult to extract general quantitative conclusions from the
published data, because of the limited number of models and the variety
of employed microphysical treatments and numerical setups. Moreover,
most of the published waveforms are not sampled at a
sufficiently high rate to capture all of the relevant features of the
signal, particularly after the first few hundred milliseconds after core
bounce, and/or were obtained from simulations that treated the inner
core of the \ac{PNS} in 1D, possibly affecting the development of the
inner \ac{PNS} convection \citep{Buras:2005tb, Dessart:2005ck,
Radice:2017ykv, Glas:2018vcs}.

In this \emph{Letter}, we report on the \ac{GW} signal from eight, new,
3D, state-of-the-art neutrino-radiation hydrodynamics \ac{CCSN}
simulations performed with the Eulerian radiation-hydrodynamics code
\fornax{} \citep{Skinner:2015uhw, Skinner:2018iti}. We present
well-sampled \ac{GW} waveforms and study, for the first time, their
generic properties using a homogeneous set of simulations covering a
wide range of \ac{ZAMS} masses and post-bounce dynamics. We show that
\ac{GW} observations could constrain the structure of the \ac{PNS} and
the magnitude of the turbulent energy fluxes impinging on it.

\section{Methods.}
We consider seven stellar evolution progenitors from
\citet{Sukhbold:2015wba} with \ac{ZAMS} masses of $9 M_\odot$, $10
M_\odot$, $11 M_\odot$, $12 M_\odot$, $13 M_\odot$, $19 M_\odot$, and
$60 M_\odot$. We also consider the $25 M_\odot$ progenitor from
\citet{Sukhbold:2017cnt}. All models have solar metallicity. We
simulate the collapse of each progenitor in 1D until 10~ms
after core bounce, since large-scale deviations from spherical
symmetry are not expected for nonrotating progenitors during the
collapse phase.  Afterwards, we remap fluid and neutrino-radiation
quantities to 3D, and we add small, dynamically unimportant, velocity
perturbations to break the spherical symmetry. In particular, we perturb
the velocity in the region $200\ {\rm km} \leq r \leq 1,\!000\ {\rm km}$
using the prescription introduced by \citet{Mueller:2014oia} with a
maximum amplitude perturbation of $100\ {\rm km}\ {\rm s}^{-1}$.
The perturbations amount to a less than 0.5\% change in the
velocity field and are expected to be dynamically irrelevant.

The evolution is continued on a grid of $678 \times 128 \times 256$
zones in $(r,\theta,\phi)$ extending up to $20,\!000\ {\rm km}$. The
radial grid is linearly spaced in the inner ${\sim}20$~km, and
logarithmic outside. These are among the highest resolution 3D
full-physics \ac{CCSN} simulations to date. Because of the extreme
computational costs, we are not able to perform a resolution study of
our results. However, because of the turbulent nature of \acp{CCSN} and
on the basis of previous studies employed simplified microphysics
\citep{Hanke:2011jf, Takiwaki:2013cqa, Abdikamalov:2014oba,
Radice:2015qva}, we expect that the detailed quantitative evolution of
each model will be stochastic. For this reason, in our analysis we will
focus on features that are found to be present in all our models and are
expected to be robust. Our grid is derefined in angle as needed to keep
the aspect ratio of the cells roughly constant when approaching the grid
center or the axis \citep{Skinner:2018iti}. This allows us to evolve the
collapsing core of the star in 3D all the way to the center.

Stellar and nuclear matter is treated using the SFHo equation of state
(EOS; \citealt{Steiner:2012rk}) \acused{EOS}.  We assume nuclear
statistical equilibrium to hold everywhere in our computational domain.
Neutrino radiation is treated using a multi-dimensional moment method
with analytical closure.

We employ twelve logarithmically spaced energy bins for $\nu_e$ and
$\bar{\nu}_e$, while heavy-lepton neutrinos are lumped together into a
single effective species ``$\nu_\mu$.'' Our neutrino treatment accounts
for gravitational redshift, Doppler effects, and inelastic
scattering~\citep{Burrows:2016ohd, Vartanyan:2018iah}.  Together with
our previous calculations \cite{Vartanyan:2018iah}, these are the only
simulations including neutrino-matter inelastic scattering in the
context of a truly multi-dimensional neutrino-transport scheme.  In
particular, we do not use the ray-by-ray method \citep{Skinner:2015uhw,
Glas:2018oyz}.

Gravity is treated in the monopole approximation using an effective
general-relativistic potential \citep{2006A&A...445..273M}. \acp{GW} are
estimated using the quadrupole approximation \citep{Finn:1990qf}, and
evaluated at every timestep ${\sim}10^{-6}\ {\rm s}$. For the analysis,
we downsample to 16,384~Hz, the data readout frequency of Advanced LIGO
\citep{TheLIGOScientific:2014jea}.

\section{Results.}
\begin{figure}
  \includegraphics[width=\columnwidth]{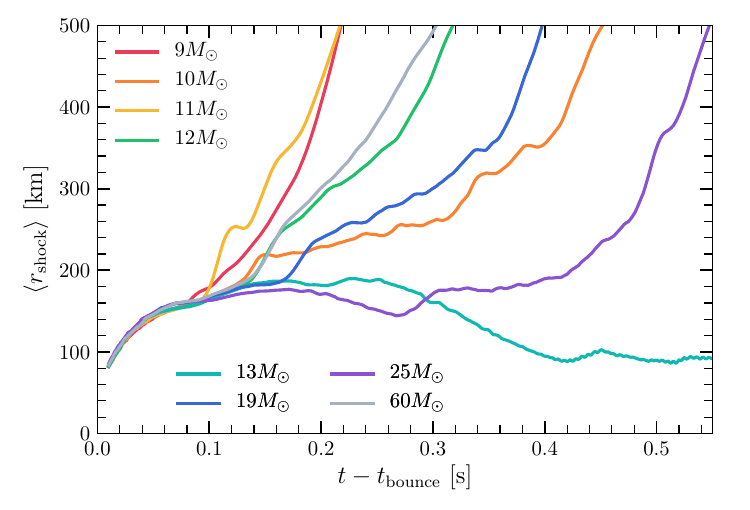}
  \caption{Average shock radius as a function of time from bounce. All
  models apart from the $13 M_\odot$ model successfully explode. The
  explosion times vary between ${\sim}0.2\ {\rm s}$ and ${\sim}0.5\ {\rm
  s}$ after core bounce.}
  \label{fig:rshock}
\end{figure}

Runaway shock expansion occurs for all but one of our progenitors (see
Fig~\ref{fig:rshock}). The explosions proceed in accordance with the
general expectations from the delayed neutrino mechanism
\citep{Colgate:1966ax, Burrows:1993pi}. The inclusion of inelastic
scattering and many-body corrections to neutrino-matter cross sections,
and the presence of sharp compositional interfaces in most of the
progenitors we considered are key for the successful explosion we
witness in our calculations \citep{Burrows:2016ohd, Vartanyan:2018iah}.
Even after shock runaway, asymmetric accretion onto the \ac{PNS}
persists for most of our progenitors to late times. The only exception
is the $9 M_\odot$ progenitor, for which the shock runaway is followed
by the emergence of an almost isotropic neutrino-driven wind. This
completely terminates accretion onto the PNS for this model. We
documented the same behavior in previously published simulations of the
same progenitor in 2D \citep{Radice:2017ykv} and in 3D
\citep{Burrows:2019rtd}.

\begin{figure}
  \includegraphics[width=\columnwidth]{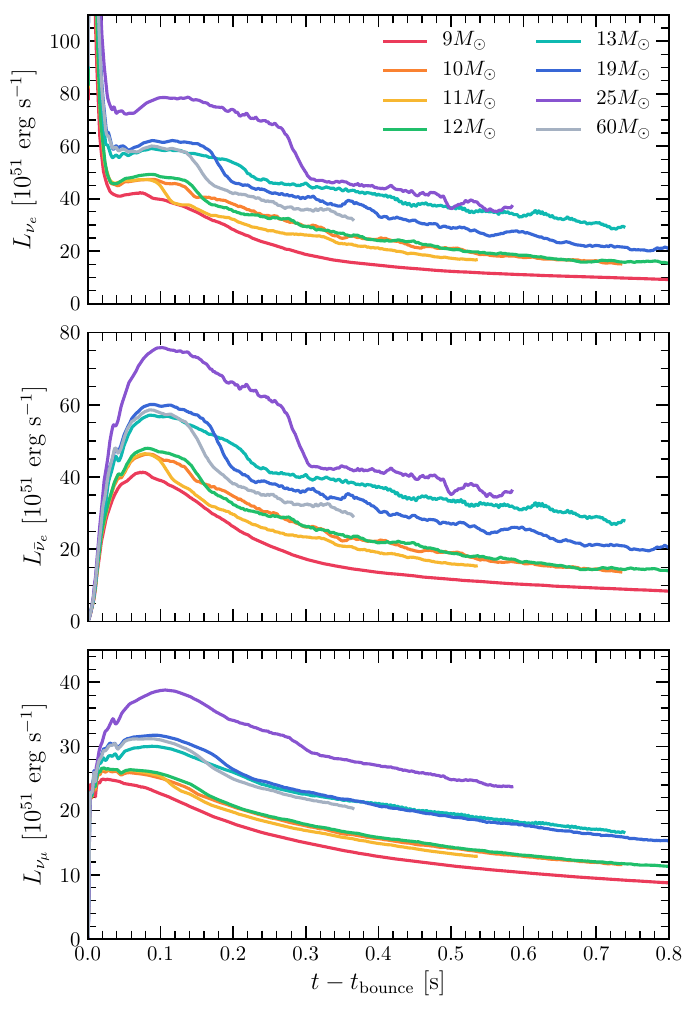}
  \caption{Neutrino luminosities at 10,000~km as a function of retarded
  time. The sudden drop in the electron-type neutrino luminosity
  experienced by some progenitors correspond to the accretion of
  compositional interfaces. Our model span a wide range in neutrino
  luminosity ranging from that of the $9 M_\odot$ progenitor to that of
  the $25 M_\odot$ progenitor.}
  \label{fig:lnu}
\end{figure}

The neutrino luminosities from our simulations are collected in
Fig.~\ref{fig:lnu}. They are bounded from below by the luminosity of the
$9 M_\odot$ progenitor, and from above by the luminosity of the $25
M_\odot$ progenitor. High neutrino luminosities are characteristic of
progenitors with higher compactnesses and accretion rates. These, in
turn, increase with the ZAMS mass for most of the progenitors we
consider here. For this reason, we find that the neutrino luminosity
increases with ZAMS mass. The exception is the $60 M_\odot$ progenitor.
This progenitor shed a significant fraction of its mass to stellar winds
and has a less compact core than the $19 M_\odot$ progenitor at the time
of collapse. Overall, this figure demonstrates the wide variety of the
progenitors considered in this work. A more detailed account of our new
calculations is presented in \citet{Burrows:2019rtd} and in Radice et
al. (2019) in prep. See also \citet{Skinner:2018iti} and
\citet{Vartanyan:2018iah} for additional information about our new set
of 3D simulations. Here, we focus only on the \ac{GW} signal from these
models.

\begin{figure*}
  \includegraphics[width=\textwidth]{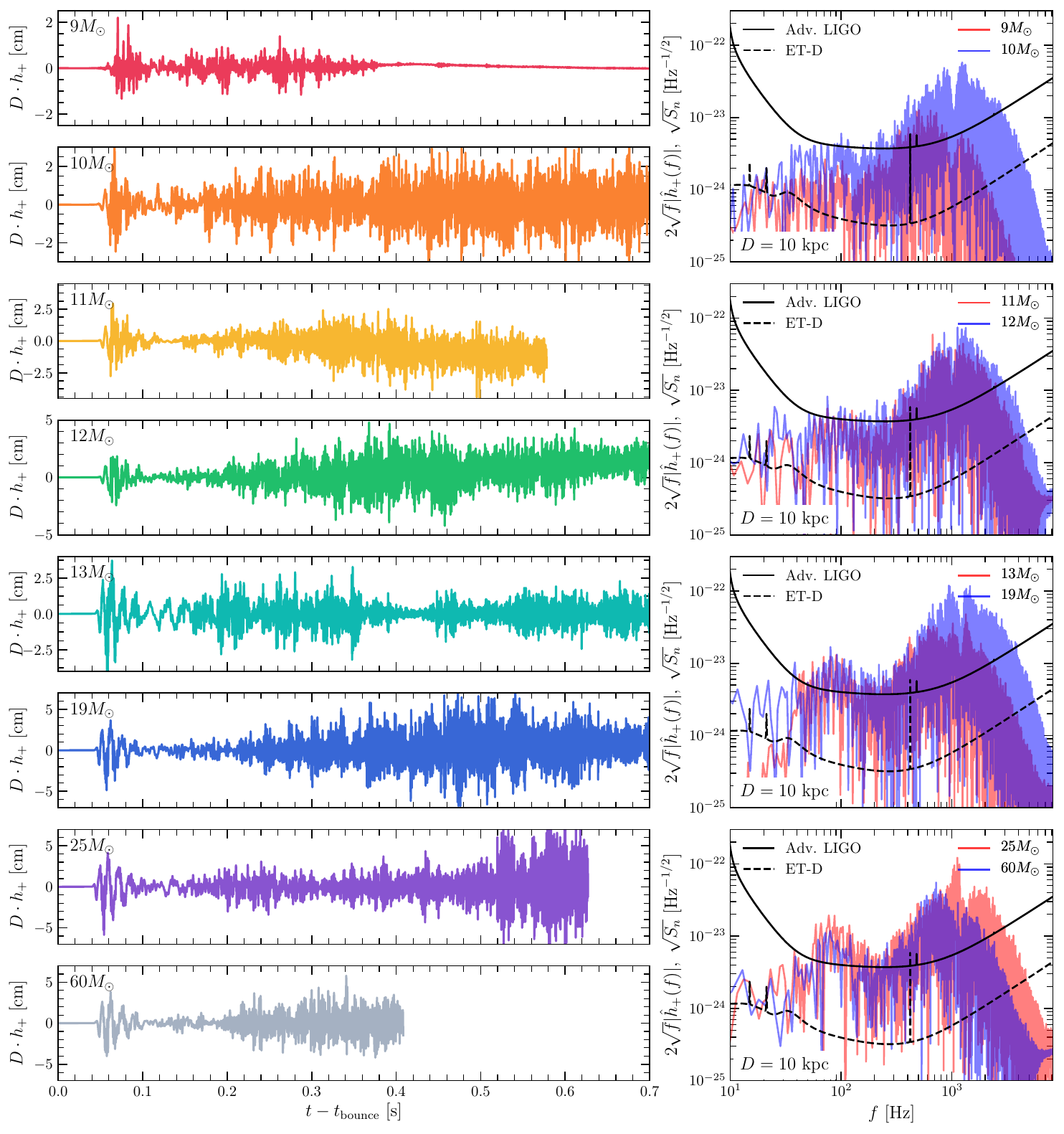}
  \caption{Plus polarization of the GW strain, and spectra from our
  models. The data is shown for an observer located along the $x-$axis.
  GW emission starts shortly after we map our models from 1D to 3D due
  to the development of prompt convection just after neutrino shock
  breakout. This early time component dominates the low frequency
  ${\sim}100\ {\rm Hz}$ part of the spectrum if SASI is absent.
  Otherwise both SASI and prompt convection contribute signal in this
  frequencies, albeit at different times (\emph{cf}.,
  Fig.~\ref{fig:specgram.s25}). After a brief quiescent phase, the GW
  amplitude starts growing again as accretion plumes perturb the
  protoneutron star. This latter part of the signal increases in
  frequency over time and determines the signal at frequencies between
  several hundred Hz and few kHz.}
  \label{fig:hp}
\end{figure*}

The \ac{GW} strains from our models are shown in Fig.~\ref{fig:hp}.
As in previous studies, we find that the \ac{GW} signal starts with a
burst shortly after bounce. This is due to prompt convective overturn
developing in conjunction with neutrino shock breakout
\citep{1987ApJ...318L..57B, Murphy:2009dx, Mueller:2012sv, Ott:2012mr,
Yakunin:2015wra}. The initial \ac{GW} burst is followed by a ${\sim}100\
{\rm ms}$ phase of quiescence that ends when neutrino-driven convection
\citep{Burrows:1995ww, Janka:1996tu, Foglizzo:2005xr, Radice:2015qva,
Radice:2017kmj}, or the standing accretion shock instability
(\acused{SASI}\ac{SASI}; \cite{Blondin:2002sm, Foglizzo:2006fu,
Burrows:2012yk, Hanke:2013jat, Abdikamalov:2014oba}) become fully
developed. Subsequently, the \ac{GW} emission is sustained by
non-spherical, intermittent accretion streams hitting the \ac{PNS} and
exciting its quadrupolar oscillation modes \citep{Murphy:2009dx,
Mueller:2012sv, Fuller:2015lpa, Morozova:2018glm, Torres-Forne:2017xhv,
Torres-Forne:2018nzj}.

\begin{figure}
  \includegraphics[width=\columnwidth]{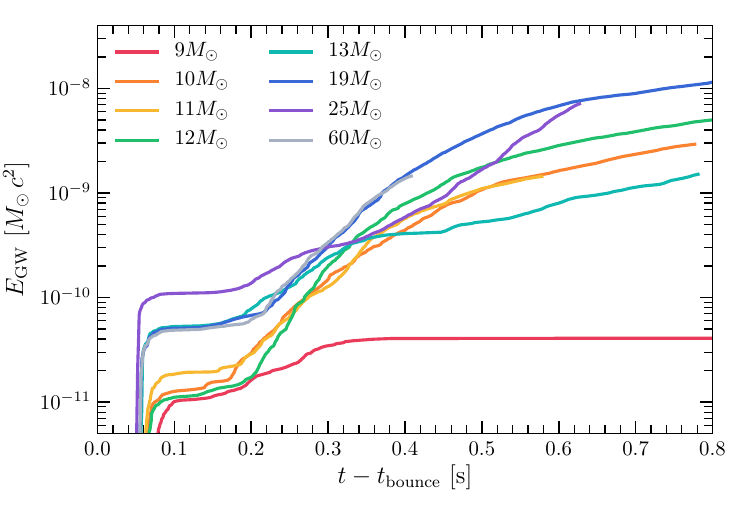}
  \caption{Integrated GW luminosity as a function of time. The radiated
  GW energy is still growing at the end of our simulations, with the
  exception of the $9 M_\odot$ progenitor which saturates at $t - t_{\rm
  bounce} \simeq 0.3\ {\rm s}$. We find that up to several times $\times
  10^{-9} M_\odot c^2$ of energy are radiated in GWs in the first half
  second after bounce.}
  \label{fig:egw}
\end{figure}

The energy radiated in \acp{GW} is shown in Fig.~\ref{fig:egw}. Our most
optimistic models emit up to several times $10^{-9} M_\odot c^2$ in the
first half second after bounce, in good agreement with the model
considered by \citet{Yakunin:2017tus}. The corresponding optimal,
single detector \acp{SNR} for Adv.~LIGO, \ie, the SNR computed
assuming perfect knowledge of the waveform, at 10~kpc range from
${\sim}1.5$ for the $9 M_\odot$ progenitor, for which the signal shuts
down at $t - t_{\rm bounce} \simeq 0.3\ {\rm s}$, to ${\sim}11.5$ for
the $19 M_\odot$ progenitor, which remains a loud \ac{GW} emitter for
the entire duration of our simulation. For the proposed \ac{ET} in the
``D'' configuration \citep{Punturo:2010zza, Hild:2010id}, which we take
as a representative 3rd-generation detector, the corresponding \acp{SNR}
are ${\sim}20$ and ${\sim}110$. These values are similar to those
reported by \citet{Andresen:2016pdt} and \citet{Andresen:2018aom} for their
models.  They imply that, even though there are good prospects for the
detection of nearby \acp{CCSN} with current generation \ac{GW}
observatories, 3rd-generation detector sensitivities are required for
confident, high-\ac{SNR} detection of all \ac{CCSN} events in the Milky
Way.

\begin{figure}
  \includegraphics[width=\columnwidth]{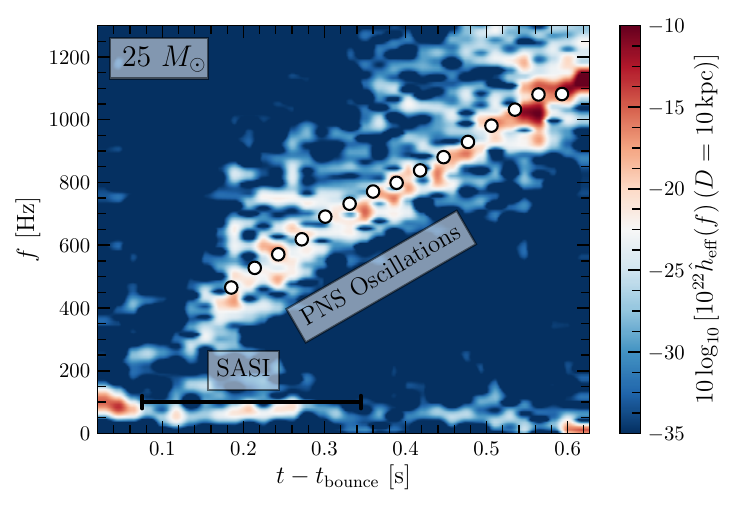}
  \caption{Time-frequency content of the GW signal for the $25 M_\odot$
  progenitor. The white dots denote the eigenfrequencies associated with
  the quadrupolar $f$- and low-order, $n = 1, 2$, $g$-modes of the PNS
  as computed from linear perturbation theory. This progenitor is the
  only one in our set showing a clear signature of the SASI at
  low-frequency. The presence of a higher-frequency component associated
  with PNS oscillations is instead universal.}
  \label{fig:specgram.s25}
\end{figure}

All the \ac{GW} waveforms from our models are characterized by the
presence of a narrow track in the time-frequency plane with steadily
increasing frequency. We show a representative example of this feature
in Fig.~\ref{fig:specgram.s25}. Using the astro-seismological approach
we developed in \citet{Morozova:2018glm}, we identify this feature with a
low-order, quadrupolar surface $g$-mode of the \ac{PNS}.  This mode
evolves as the \ac{PNS} contracts, increasing in frequency, and assuming
the character of a quadrupolar $f$-mode when $t - t_{\rm bounce} \gtrsim
0.4\ {\rm s}$. We observed an identical trend in our previous 2D study
\citep{Morozova:2018glm}. This is expected, since \ac{PNS} masses
and radii found in our 3D simulations are in excellent agreement with
those found in the corresponding 2D simulations.

A lower frequency feature of the \ac{GW} signal associated with the
\ac{SASI}, is present only for the $25 M_\odot$ progenitor. The $13
M_\odot$ progenitor also shows \ac{SASI} activity at late times, but
this is not accompanied by a strong \ac{GW} signal. This might be due to
the fact that the accretion rate for the $13 M_\odot$ model is smaller
than that of the $25 M_\odot$ model, which implies that a smaller amount
of material is involved in the \ac{SASI} motion for the former. The time
interval over which the \ac{SASI} is active for the $25 M_\odot$
progenitor, as well as the associated characteristic frequency, are
highlighted in Fig.~\ref{fig:specgram.s25}. \citet{Hayama:2016kmv} and
\citet{Hayama:2018zgb} reported that that rotation and/or \ac{SASI}
activity could leave an imprint in the circular polarization of the
\ac{GW} signal. However, we do not find evidence for this effect in our
simulations. This is possibly because \ac{SASI} is not as vigorous in
our models as it is in theirs. The \ac{SASI} signal disappears once
runaway shock expansion develops, in agreement with previous findings
\citep{Andresen:2016pdt, Andresen:2018aom}. Apart from the disappearance
of the \ac{SASI} signature for the $25 M_\odot$ progenitor and the
vanishing of the \ac{GW} emission from the $9 M_\odot$ model, we do not
find obvious signatures of explosion, or lack thereof, in the \ac{GW}
signals.

It has been speculated that \ac{PNS} convection might be the main agent
perturbing the \ac{PNS} and driving the emission of \acp{GW}
\citep{Andresen:2016pdt}. However, our $9 M_\odot$ progenitor seems to
rule out this hypothesis: the \ac{PNS} convection for this model is
vigorous throughout the evolution, but the \ac{GW} luminosity decays
substantially after $t - t_{\rm bounce} \simeq 0.3\ {\rm s}$ (see
Fig.~\ref{fig:egw}). Instead, the drop in the \ac{GW} luminosity for
this model is coincident with the emergence of a quasi-spherical wind
from the \ac{PNS} and the termination of accretion. This suggests
instead that it is the chaotic accretion onto the \ac{PNS} that is
driving the \ac{GW} emission.

\begin{figure}
  \includegraphics[width=\columnwidth]{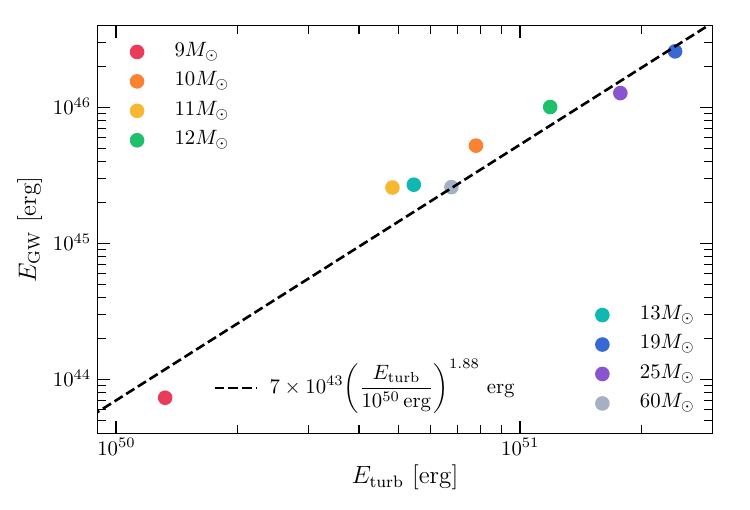}
  \caption{Energy radiated in GWs versus time-integrated power of the
  turbulent flow impinging upon the PNS. Note that, for most of our
  models, these values are still growing at the end of our simulations
  (see Fig.~\ref{fig:egw}). $E_{\rm GW}$ and $E_{\rm turb}$ are
  strongly correlated suggesting that \ac{GW} observations could
  constrain the strength of the turbulence behind the shock.}
  \label{fig:Egw_vs_Eturb}
\end{figure}

To test this hypothesis we compute time-integrated turbulent energy
fluxes (kinetic plus thermal) impinging on the \ac{PNS} using to the
formalism derived in \citet{Radice:2015qva}. We then compare the total
amount of turbulent energy accreted by the \ac{PNS}, $E_{\rm turb}$, to
the total amount of energy irradiated in \acp{GW}, $E_{\rm GW}$. The
results are given in Fig.~\ref{fig:Egw_vs_Eturb}. We find a clear
correlation between $E_{\rm GW}$ and $E_{\rm turb}$. This is evidence
for accretion being the main driver of the \ac{GW} emission.
Specifically, our results suggest that \acp{GW} are produced by the
non-resonant excitation of pulsational modes of the \ac{PNS} by chaotic
accretion. We remark that a scaling close to $E_{\rm GW} \sim
E_{\rm turb}^2$ can be expected on the basis of simple dimensional
arguments \citep[][chapter 36]{1973grav.book.....M} essentially because
of the quadrupolar nature of \acp{GW} \citep{Muller:2017vuu}.

\section{Discussion.}
We have analyzed the \ac{GW} signals from a large set of 3D \fornax{}
\ac{CCSN} simulations. Our calculations employed the most advanced
treatment for neutrino transport and neutrino-matter interactions
available. The most robust feature is an excess in the time-frequency
diagram of the \ac{GW} strain following a characteristic track. The
corresponding peak frequency is associated with quadrupolar oscillation
modes of the \ac{PNS}, so its measurement would allow us to constrain
the structure of the \ac{PNS}. This in turn would have consequences for
the \ac{EOS} and for the transport properties of warm nuclear matter.

A signature of the \ac{SASI} is found only in one progenitor. If present
and detected, this signature could potentially be extremely valuable
because it might be used to infer the time at which the supernova shock
is revived \citep{Andresen:2016pdt, Andresen:2018aom}. In combination
with the knowledge of the time of neutrino shock breakout
\citep{Wallace:2015xma}, this would produce a strong constraint on the
explosion mechanism.  Nevertheless, our results show that, in contrast
with what \citet{Andresen:2016pdt} and \citet{Andresen:2018aom} claimed
on the basis of a few models affected by aliasing, it is the pulsation
of the \ac{PNS} that is the most robust feature of the \ac{GW} signal in
3D, and not the \ac{SASI}. Indeed, a signature of the \ac{SASI} is found
only in one of our progenitors.

Our simulations also clearly demonstrate that \acp{GW} are driven by
convection at the periphery of the \ac{PNS} and not by the convection
inside the \ac{NS} as claimed by \citet{Andresen:2016pdt}. Finally, we
have shown, for the first time, that a measurement of the overall
amplitude of the \ac{GW} signal would constrain the strength of
turbulence induced by neutrino-driven convection or \ac{SASI} behind the
shock. This would allow us to probe directly the engine of \acp{CCSN}.
Our results show that \ac{GW} observations are a promising avenue by
which to probe the otherwise inaccessible dynamics of the inner engine
of \acp{CCSN}. However, these observations will likely require the kind
of high sensitivity over a broad range of frequencies that only future
generation \ac{GW} detectors can achieve.

Future work should develop the data analysis techniques necessary to
extract the features we have identified in the waveforms, as well as
systematic strategies to jointly analyze \ac{GW} and neutrino signals.
First, the detection of the neutrino burst will reveal the time and sky
position of the \ac{SN}, thus reducing the false alarm rate for the
\ac{GW} signal and the number of free parameters needed for
template-based searches \citep{Adams:2013ana, Nakamura:2016kkl}. Second,
the detection of correlated neutrino and \ac{GW} temporal variability
might provide a way to diagnose large-scale chaotic motion in the
supernova core \citep{Ott:2012kr, Kuroda:2017trn}. We also plan to extend
this work with the study of progenitors with moderate rotation and with
relic perturbations from advanced nuclear burning stages
\citep{Couch:2015gua, Muller:2017hht}, and to explore the \ac{GW} signal
over longer timescales.

\acknowledgements
We acknowledge support via the Scientific Discovery through Advanced
Computing (SciDAC4) program and Grant DE-SC0018297 (subaward 00009650),
the U.S. NSF under Grants AST-1714267 and PHY-1144374, partial support
for DR as a Frank and Peggy Taplin Fellow at the Institute for Advanced
Study, and the allocation of generous computer resources under the NSF
PRAC program at Blue Waters (award \#OAC-1809073), under XSEDE at
Stampede2 (ACI-1548562), and at NERSC under their contract
DE-AC03-76SF00098.

\software{
  \fornax{} \citep{Skinner:2018iti},
  NumPy \citep{numpy},
  Matplotlib \citep{matplotlib}
}



\end{document}